%% file: hdfs_spec.tex
\documentclass[twocolumn]{aa}
\usepackage{graphicx}
\def\lesssim{\mathrel{\hbox{\rlap{\hbox{\lower4pt\hbox{$\sim$}}}\hbox{$<$}}}}
\def\gtrsim{\mathrel{\hbox{\rlap{\hbox{\lower4pt\hbox{$\sim$}}}\hbox{$>$}}}}
\begin{document}
\title{Spectroscopy and Stellar Populations of Star-forming 
Galaxies at $z \sim 3$ in the Hubble Deep Field - South
\thanks{Based on observations collected at the European 
Southern Observatory, Paranal (Chile); Proposal No.: 
71.A-0419(A).}}

   \author{I. Iwata,\inst{1,2,3}
          A. K. Inoue,\inst{4,3,5} \fnmsep\thanks{JSPS Postdoctral 
           Fellow for Research Abroad}
           \and
D. Burgarella\inst{4}
          }

   \offprints{I. Iwata}

\institute{Okayama Astrophysical Observatory, National Astronomical Observatory 
of Japan,\\ Honjo, Kamogata, Okayama 719-0232 Japan
              \email{iwata@oao.nao.ac.jp}
\and
Subaru Mitaka Office (Subaru Telescope), National Astronomical Observatory 
of Japan,\\ Osawa 2-21-1, Mitaka, Tokyo 181-8588 Japan
\and
Department of Astronomy, Faculty of Science, 
Kyoto University, Kitashirakawa-Oiwakecho, Sakyo-ku, Kyoto 606-8502 Japan
\and
Laboratoire d'Astrophysique de Marseille, Traverse du
Siphon, BP 8, 13376 Marseille, CEDEX 12, France
\and
Department of Physics, Faculty of Science, 
Kyoto University, Kitashirakawa-Oiwakecho, Sakyo-ku, Kyoto 606-8502 Japan
}

   \date{Received February 17,2005; accepted May 24, 2005}

\abstract{
We present results of VLT/FORS2 spectroscopy of galaxies at $z \sim 3$ 
in the Hubble 
Deep Field - South (HDF-S). A sample of galaxies was drawn from the 
photometric redshift catalogue based on the HST/WFPC2 optical images and 
the deep near-infrared images obtained with VLT/ISAAC as a part of 
the Faint Infrared Extragalactic Survey (FIRES) project. 
We selected galaxies with photometric redshift 
between 2.5 and 4. Most of the selected galaxies are bright in rest-frame 
UV wavelengths and satisfy color selection criteria of 
Lyman break galaxies (LBGs) at $z \sim 3$. 
The number of target galaxies with $I_\mathrm{AB} \leq 25.0$ was 15.
We identified new 5 firm and 2 probable redshifts in addition to confirmations 
of previously known 6 galaxies at $z \sim 3$. We found 6 among these 13 
galaxies lie at a quite narrow redshift range at $z = 2.80 \pm 0.01$. 
Their spatial distribution is 
fairly concentrated and is at the edge of the HDF-S field, suggesting the 
possible existence of larger galaxy clustering. 
We examined stellar populations of the galaxies with spectroscopic redshifts 
through comparisons of their 
optical and near-infrared photometry data with template spectra generated by 
a population synthesis code.
The ages from the onset of star formation for these star-forming galaxies 
with $I \leq 25.0$ are typically 50--200 Myr, and their stellar masses 
are between (0.5--5) $\times 10^{10} M_{\sun}$, consistent with previous studies.
We also compared these SED fitting results with those for ``distant red galaxies'' 
(DRGs) at $z > 2$ discovered by FIRES. 
DRGs have larger stellar masses, larger dust attenuation than 
our UV-luminous LBG sample, and their star formation rates are 
often comparable to LBGs. 
These trends suggest that majority of DRGs are indeed the most massive 
systems at the redshift and are still in the active star-forming phase. 
Unless the number density of DRGs is much smaller than LBGs, 
estimates based on UV selected sample 
could miss substantial part of stellar mass density at $z \sim 3$. 
}

\titlerunning{Galaxies at $z \sim 3$ in the HDF-S}
\authorrunning{I. Iwata et al.}
\maketitle

%

\section{Introduction}

In recent several years, there has been much progress in understanding 
the nature of galaxies at high-redshift. Deep imaging of blank fields and 
the photometric redshift technique based on multi-band photometry 
have enabled a construction of 
large samples of high-redshift galaxies which are so faint that we need long 
exposures to obtain good signal-to-noise spectroscopy of them. 
The Lyman break technique, a simple variant of the photometric redshift techniques, 
intends to detect high redshift galaxies 
by finding a spectral break at redshifted Lyman limit and the discontinuity 
at shortward of Ly$\alpha$ due to attenuation by intergalactic neutral 
hydrogen. 
The number of galaxies selected by the technique, so-called Lyman break galaxies 
(LBGs), amounts to an order of $10^3$ at $z \sim 3$ 
(e.g., Steidel et al. \cite{ste96}, \cite{ste03}; 
Foucaud et al. \cite{fou03}), and several hundreds even at $z \sim 5$ 
(Iwata et al. \cite{iwt03}; Ouchi et al. \cite{ouc04}). 
Follow-up studies on LBGs such as optical spectroscopy 
(e.g., Teplitz et al. \cite{tep00}; Pettini et al. \cite{pettini01}; 
Shapley et al. \cite{shap03}; Ando et al. \cite{and04}) 
and analysis of their stellar populations using optical-to-near-infrared 
spectral energy distribution (SED) (e.g., Sawicki and Yee \cite{sy98}; 
Shapley et al. \cite{shap01}; Papovich et al. \cite{pap01}) 
have been made. By these studies basic properties of LBGs have been 
clarified; their star-formation rates amount to several tens to 
several hundreds $M_{\sun}$/year, dust attenuation is fairly small 
($E(B-V)<0.5$), their stellar populations are dominated by 
relatively young massive stars (ages less than 1 Gyr in most cases) 
and stellar mass is an order of $10^{10} M_{\sun}$ 
(about 1/10 of the Galaxy). 

Although these studies have clearly indicated that 
the Lyman break technique is a powerful method to probe the 
star-forming galaxies at $z \gtrsim 3$, the number fraction of 
LBGs among the total populations of galaxies at a redshift considered 
remains unclear.
Since the Lyman break technique selects galaxies 
which show prominent discontinuity at $\lambda = 1216$\AA\ and 
have flat UV spectra, sample galaxies are biased to actively 
star-forming galaxies with relatively small amount of dust attenuation. 
Recently, several studies based on deep near-infrared imaging 
revealed substantial populations of massive galaxies which 
are faint in optical bands but bright in near-infrared 
(e.g., Franx et al. \cite{frx03}; Daddi et al. \cite{dad04}; 
Chen and Marzke \cite{chen04}), and it has been suggested that 
optical (i.e., rest-frame UV) based samples may miss a significant 
amount of massive galaxies. 
However, due to their optical faintness, only a small fraction of 
these red galaxies have spectroscopic redshifts. 
Thus current estimates of their redshift distribution and number 
density have large uncertainties. 
Moreover, since the red color can be 
explained by both heavy dust attenuation and old stellar population, 
the nature of these galaxies -- dusty star-forming galaxies or 
massive galaxies dominated by old stellar population -- is 
unresolved. 

The Hubble Deep Fields (HDFs) are one of the most prominent observations for 
the study of high-z galaxies. Deep imaging covering the optical wavelengths 
by HST/WFPC2 (Williams et al. \cite{wil96}; Casertano et al. \cite{cas00}) 
as well as near-infrared images 
obtained by HST/NICMOS and ground-based observatories enabled the 
photometric redshift estimation down to $I_\mathrm{AB}=27.5$, 
for both northern and southern HDFs. 
(e.g., Fern\`{a}ndez-Soto et al. \cite{fsoto99}; Yahata et al. \cite{yah00}; 
Labb\'{e} et al. \cite{lab03} (hereafter L03); 
Vanzella et al. \cite{van04}). 
In the HDF-S we carried out a deep OII+44 narrow-band imaging 
using VLT/FORS1 to examine the escape fraction of 
Lyman continuum from galaxies at $z \sim 3$ (Inoue et al. \cite{ino05}). 
Although precise redshift measurements were required to use the narrow-band 
image to discuss the Lyman continuum escape fraction, 
in the HDF-S only a handful spectroscopic observations have been made so far 
(Cristiani et al. \cite{cri00}; 
Vanzella et al. \cite{van02}; Sawicki and Mall\'{e}n-Ornelas \cite{saw03}; 
Rigopoulou et al. \cite{rig05}), and in particular, the number of 
galaxies identified to be at $z > 2.5$ is still quite limited; 
only 2 galaxies at $z > 3$ and 8 galaxies at $2.5 < z < 3$ have been 
reported. In order to increase the number of sample galaxies, 
we carried out spectroscopic observations of galaxies with photometric 
redshifts estimated to be at $z \sim 3$ with VLT/FORS2.
The precise redshift information is also 
useful to reduce the uncertainty of physical properties 
(stellar mass, ages, amount of dust attenuation) in SED fitting.
In this paper we report the results of the spectroscopic observations. 
In section 2 we describe the sample selection, observations and 
data reduction. The results and a discussion on a spatial distribution 
of sample galaxies are presented in section 3. 
We also investigate the stellar population and star formation history 
of the galaxies (section 4), and compare them with 
properties of ``distant red galaxies'' discovered by Franx et al. 
(\cite{frx03}) in section 5. A summary and conclusions of the paper 
are presented in section 6. 

Throughout the paper all magnitudes are presented in AB system. 
We adopt a set of cosmological parameters of 
($\Omega_{\mathrm M}$, $\Omega_\Lambda$, $h$) = (0.3, 0.7, 0.7).


\section{Observations and data reduction}

\subsection{Sample selection and observations}

The target field is the WFPC2 main field of the 
Hubble Deep Field - South, centered at 
R.A. = 22:32:56.2 Decl=$-60$:33:02.7. 
The WFPC2 observations, data reduction, data quality 
and object detection procedures are described in Casertano et al. (\cite{cas00}). 
Ultra-deep near-infrared $Js$, $H$, $Ks$-band imaging observations using VLT/ISAAC 
have been carried out as a part of the Faint Infrared Extragalactic Survey
(FIRES). L03 
utilized their near-infrared images 
and publicly available HST/WFPC2 optical images to estimate  
photometric redshifts for faint galaxies. 
The catalog contains 134 entries with a photometric redshift 
between 2.5 and 3.7. We selected a primary sample for spectroscopic 
observation with the $I_\mathrm{814}$ corrected isophotal magnitude brighter 
than 25.0. The number of galaxies satisfying the criteria was 23.

\begin{figure}
\centering
\resizebox{\hsize}{!}{\includegraphics{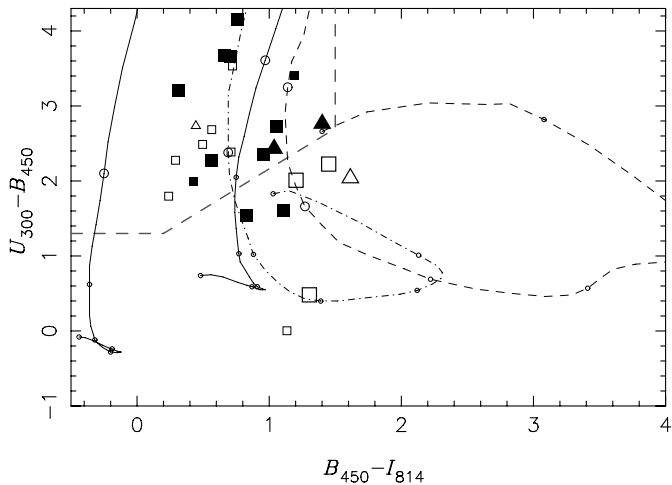}}
\caption{$U_{300}-B_{450}$ versus $B_{450}-I_{814}$ diagram for 
sample galaxies for VLT/FORS2 spectroscopy. 
The galaxies for which we successfully 
identified their redshifts are represented with filled symbols, 
and those observed but failed in redshift determination are 
shown as open symbols. Squares are for galaxies which have been 
detected in $U_{300}$-band, and triangles are for those without 
$U_{300}$-band detection. 
Larger marks and smaller ones are for galaxies with 
$I_\mathrm{814} < 25.0$ and those with 
$25.0 < I_\mathrm{814} < 25.5$, respectively. 
We also show the color tracks of model star-forming galaxies with 
and without dust attenuation as two solid lines. 
The model was computed assuming a constant star-formation rate and 
an age of 100 Myr.
The left track is for the model without dust, 
and the right one is with dust attenuation ($E(B-V)= 0.4$). 
These lines indicate how the colors of a star-forming 
galaxy change as its redshift increases. 
The dashed and dot-dashed curves are calculated 
using template spectra of elliptical and Sbc galaxies, respectively, 
from Coleman et al. (\cite{col80}). 
We put open circles at redshift interval of 0.5. Points for $z \geq 2.5$ 
were enlarged. 
The color selection criteria used by Madau et al. (\cite{mad96}) 
for selecting LBGs at $z > 2$ are indicated by a dashed line.
}
\label{fig:twocol}
\end{figure}

Figure \ref{fig:twocol} shows the distribution of sample galaxies in the 
$U_{300}-B_{450}$ versus $B_{450}-I_{814}$ two-color diagram. 
For those without $U_{300}$ detection, lower limits of $U_{300}-B_{450}$ 
colors are shown. Since the selection of the sample galaxies were not based 
on the Lyman break method but on the photometric redshift 
using multi-band imaging data, not all the sample galaxies 
satisfy the criteria for selection of $z \sim 3$ LBGs 
defined by Madau et al. (\cite{mad96}). 
All except two objects with spectroscopic redshift 
determinations (shown with filled symbols in figure \ref{fig:twocol}) 
satisfy the LBG color criteria (see \S 3). Two objects with spectroscopic 
redshifts lying at the outside of the LBG color area are HDFS0001 
and HDFS0085. These two objects locate close to the edges of the 
WFPC2 image, and photometric errors in $U_{300}$-band are estimated 
to be $\sim 0.5$mag. Thus we expect that properties of sample galaxies 
with spectroscopic redshifts are quite similar to LBGs. 

We made a multi-object spectroscopy using VLT/FORS2 in MXU mode.  
In this mode exchangeable mask unit with slitlets prepared on a focal plane 
were used. 
We defined the positions of slitlets on the HST/WFPC2 $I_\mathrm{814}$-band 
image, using the software provided by ESO. 
For some galaxies with $I_\mathrm{AB} \leq 25.0$, we could not 
set a mask due to overlappings with other target objects. 
On the other hand, we set some additional slitlets for fainter 
objects, i.e., $25.0 < I_\mathrm{AB} < 26.0$, if they did not 
conflict with slitlets for bright objects. The final sample 
for spectroscopy consists of 15 galaxies with $22.9 < I_\mathrm{AB} \leq 25.0$ 
and 10 galaxies with $25.0 < I_\mathrm{AB} \lesssim 26.0$. 
In table \ref{tab:sample} basic parameters of the sample galaxies are listed.

The observations were executed in service mode, on 7 nights during 
25th June to 28th July 2003. We used the 300V grism, which provides a 
1.6\AA\ / pixel resolution over 3300\AA -- 6600\AA. 
Slit widths were $1.7''$, and the resolving power at 5900\AA\ 
was 260. 
Exposure time for each shot was 999 sec in most cases, and 
the total on-source exposure time was 6.1 hours.

\subsection{Data reduction}

Basic data reduction steps, namely, bias subtraction (using overscan regions), 
cosmic-ray removal and flat-fielding have been made using IRAF. 
After the registration of all frames, the images were combined using 
the IRAF/IMCOMBINE task.
The sky subtraction and extraction of one-dimensional object spectra 
were made using an interactive program developed by one of us (I.I.). 
Using that program a user can make a background fitting in any wavelength 
range and preview the result of subtraction before execution.

The flux calibration and sensitivity correction were made using 
spectral data of standard stars (Feige 110, LTT 1020 and LTT 6248) taken 
during the observing runs.

\section{Results of spectroscopy}

\subsection{New redshift identifications}

\begin{figure*}
\centering
\includegraphics[width=16cm]{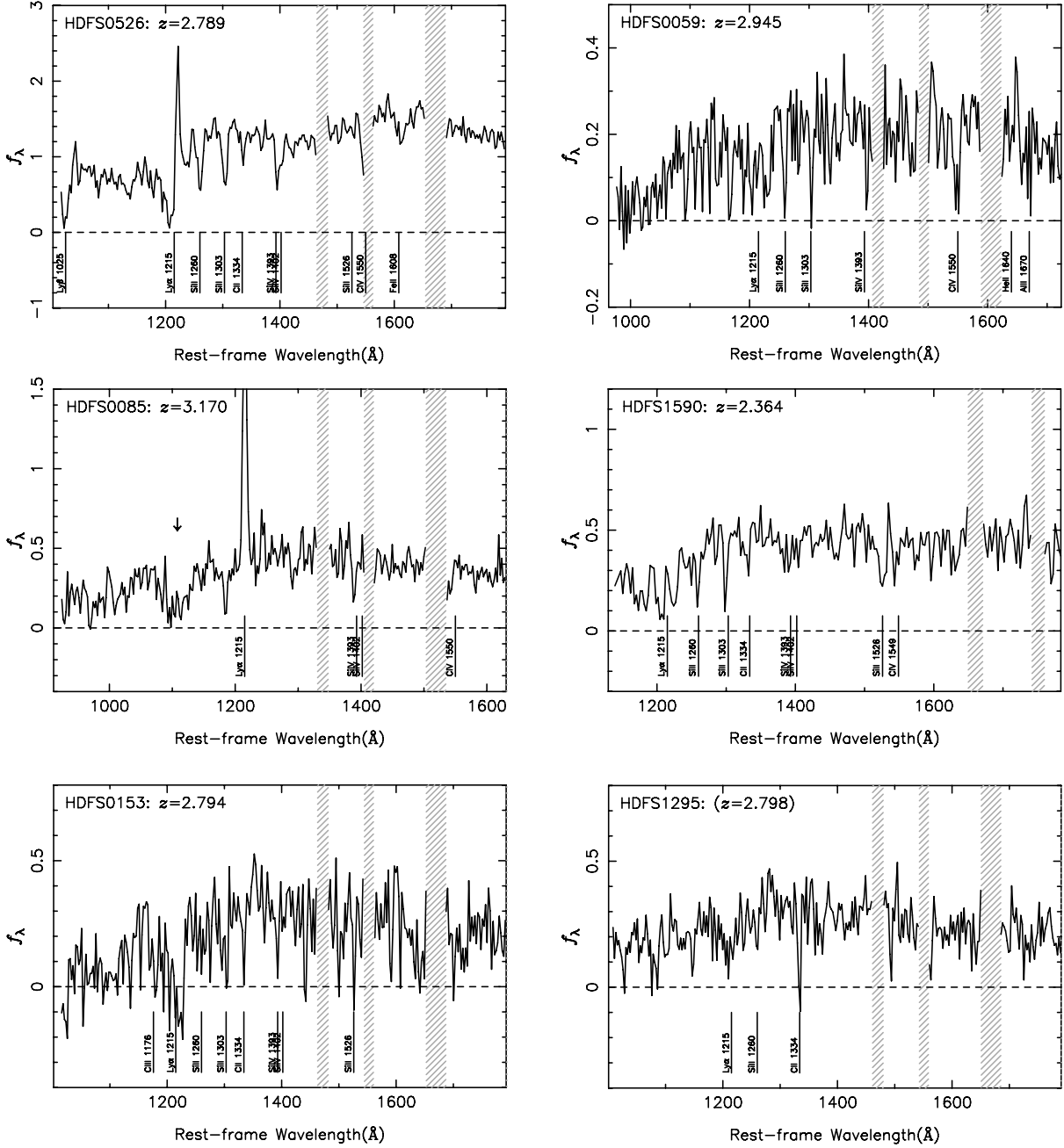}
\caption{The rest-frame UV spectra of galaxies with redshift identifications. 
Objects are shown as an order of $I_\mathrm{814}$ magnitude. 
The spectra were 3-pixel binned. Units of $f_\lambda$ is 
$10^{-16}$ erg/s/cm$^2$/\AA. 
The hatched area represent the wavelength ranges dominated by atmospheric 
lines and where we do not show objects' spectra. 
The redshifts with 
parentheses indicate that the redshift determination is not definitive. 
An arrow in the panel of HDFS0085 indicates a position of Ly$\alpha$ 
line at $z=2.8$, which coincidents with a dump in the continuum 
(see section 3.2).}
\label{fig:spectra_a}
\end{figure*}

\begin{figure*}
\centering
\includegraphics[width=16cm]{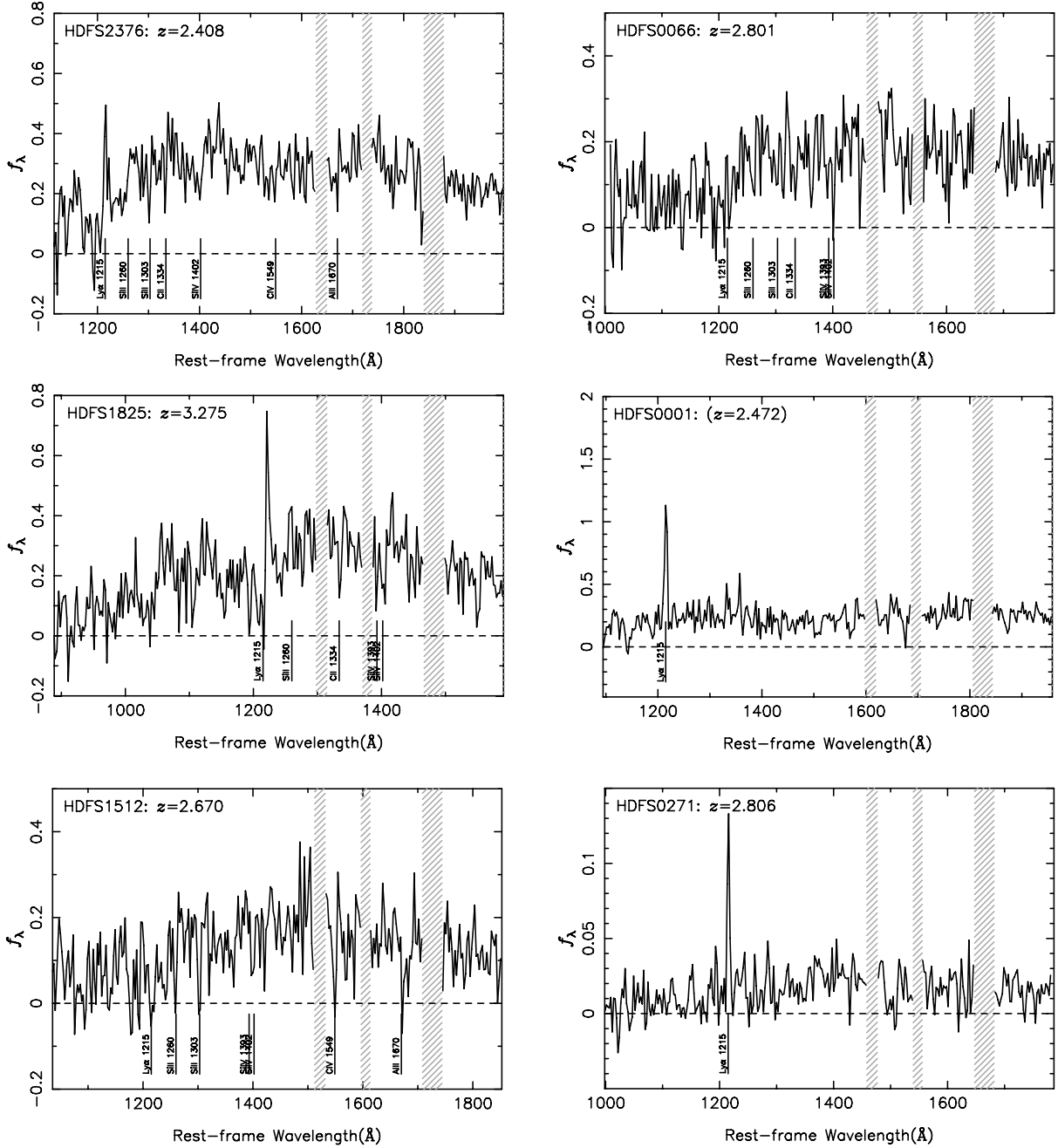}
\caption{Same as Fig. ~\ref{fig:spectra_a}, for objects fainter 
than $I_{814} \gtrsim 24.6$.}
\label{fig:spectra_b}
\end{figure*}

We firmly identified 5 new spectroscopic redshifts. 
All of these new redshift determinations are based on the identifications 
of three or more spectroscopic features, such as Ly$\alpha$ emission, 
low-ionization absorption lines (e.g., Si~{\sc ii} and C~{\sc ii}), and 
the discontinuity of continua at 1216\AA.  
They are labeled with asterisks in the sixth column of table \ref{tab:sample}.
In figure \ref{fig:spectra_a} and \ref{fig:spectra_b} we show the extracted 
one-dimensional spectra of galaxies with redshift identifications. 
There are six galaxies with previously determined redshifts; one object 
HDFS00271 is a faint ($I = 25.45$) Ly$\alpha$ emitter, and the others 
are $I < 24.4$. Redshifts for these objects listed in table \ref{tab:sample} are 
based on our spectroscopic data, and the differences from previous studies 
are less than 0.003. There is also one galaxy HDFS10652 (HDFS 223250.05$-$603356.8) 
which we did not observe 
but the redshift has been determined to be $z=2.652$ by Vanzella et al. 
(\cite{van02}).
Thus in total 12 galaxies 
with $2.5 < z < 3.5$ have been found in the HDF-S field.
There are also two objects whose redshifts are likely to be newly determined 
by our observation, but the identifications are not definite because 
the S/N of spectra was poor and only one or two lines are firmly 
identifiable. 
We list these two probable redshifts in table 
\ref{tab:sample} with parentheses. For galaxies whose spectra have 
Ly$\alpha$ line as emission lines we show the rest-frame FWHMs and 
equivalent widths (EWs) in table \ref{tab:sample}. 
For HDFS00565 and HDFS00526 we measured FWHMs and EWs for combined 
spectra, because their positions are only $2.''1$ apart and 
spectra are blended. We tried to extract spectra of their central parts 
separately and saw no difference in their redshifts ($z=2.789$).
Because the signal-to-noise ratios of spectra are poor, 
we could not measure EWs of absorption features. 

\begin{figure}
\centering
\resizebox{\hsize}{!}{\includegraphics{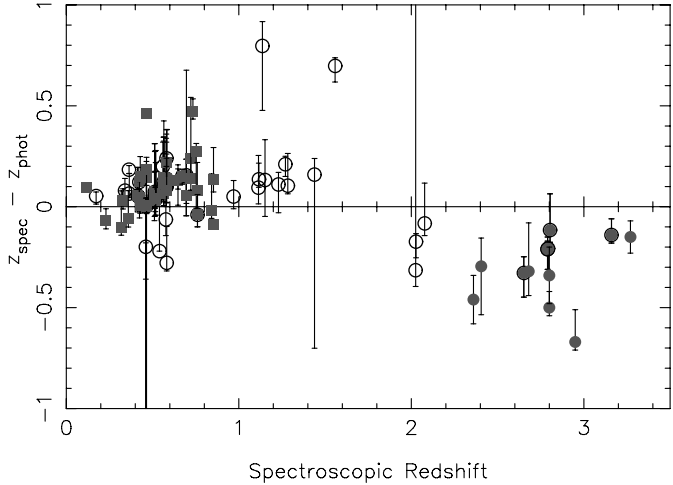}}
\caption{Comparison of Spectroscopic redshifts ($z_\mathrm{spec}$) 
and photometric redshifts ($z_\mathrm{phot}$) 
estimated by L03. 
The difference between the spectroscopic redshifts and photometric ones 
are shown along with the spectroscopic redshift. The error bars represent the 
differences ($z_\mathrm{spec} - z_\mathrm{phot}$) in cases of using 
lower and upper limits (68\% confidence level) of photometric redshifts. 
The galaxies with firm redshifts in our study are shown with filled circles. 
Open circles represent galaxies listed in L03. 
Galaxies for which $z_\mathrm{spec}$ have been measured by 
Sawicki and Mall\'{e}n-Ornelas (\cite{saw03}) 
are shown with filled squares. 
}
\label{fig:spz_phtz}
\end{figure}

The spectroscopic redshifts obtained for our 
sample galaxies are always smaller 
than photometric redshifts 
estimated by L03. 
In Fig. ~\ref{fig:spz_phtz} 
we show a comparison of spectroscopic and photometric redshifts 
for all galaxies in the HDF-S which have a published spectroscopic 
redshift. 
There is no object at $z > 2$ for which photometric redshift 
is equal to or smaller than spectroscopic redshift. On the other hand, 
the photometric redshifts of galaxies at $z < 2$ are systematically 
smaller than spectroscopic redshifts. 
This trend can be also seen in figure 6 of L03 
but it is not clear due to smaller number of galaxies with spectroscopic 
redshifts.

\subsection{Galaxy concentration at $z = 2.8$}

\begin{figure}
\centering
\resizebox{\hsize}{!}{\includegraphics{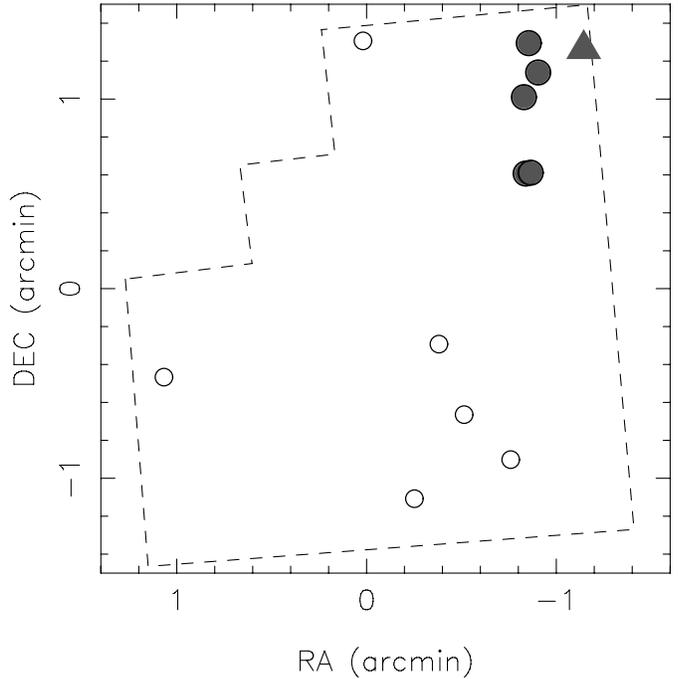}}
\caption{Spatial distribution of galaxies with 
spectroscopically confirmed redshifts. Galaxies within a 
redshift range $2.789 < z < 2.806$ are shown as filled circles. 
The other galaxies are shown as open circles except HDFS00085 at 
$z=3.16$ which is marked as a triangle. 
In the spectrum of HDFS00085 there is a signature of Ly$\alpha$ 
absorption systems 
corresponding to a HI cloud at $z\approx 2.8$. }
\label{fig:dist}
\end{figure}

We noticed that five among 12 galaxies with firm spectroscopic redshifts 
locate between a narrow redshift range from 2.789 to 2.806 
(corresponding to a comoving length of $\sim 17.6$ Mpc). 
In figure \ref{fig:dist} we show the spatial distribution of 12 galaxies 
with known redshifts. The five galaxies between $2.789 < z < 2.806$, 
shown as filled circles, also lie within a $\sim 1' \times 0.'5$ 
sky area ($1'$ corresponds to $1.8$ Mpc (comoving) at $z = 2.8$).
In addition, in the spectrum of HDFS00085 at $z = 3.16$ (shown as a 
triangle in figure \ref{fig:dist}) we identified a dump of a continuum around 
$\sim 4600$\AA, which coincides with Ly$\alpha$ clouds at $z \approx 2.80$ 
(see a dip around 1100\AA\ in the spectrum of HDFS0085 in 
figure \ref{fig:spectra_a}). This feature suggests the 
existence of large amount of neutral hydrogen at $z \sim 2.8$ toward 
the direction. We also note that one object among two with probable 
redshifts, HDFS01295 has an estimated redshift $z = 2.798$. 
It is also interesting the existence of HDFS00565 and HDFS00526. They are 
a pair of galaxies at $z=2.789$, 
the most bright galaxies in our sample galaxies in $I$-band, and 
they both show a prominent Ly$\alpha$ emission, indicating they are 
experiencing a massive star formation. 
Because the comoving volume traced by the redshift range 
$2.789 < z < 2.806$ is less than 2\% of the volume surveyed by the 
range $2.5 < z < 3.5$, the possibility that this concentration 
is caused by chance is less than $10^{-4}$ if we simply assume that 
the distribution of sample galaxies are random. 
From these findings we think that these galaxies would be members of 
a real galaxy concentration at $z \sim 2.8$ within a 30 Mpc$^3$ or 
smaller comoving volume. 
The sky positions of galaxies in the redshift range are located at the edge of 
the WFPC2 survey field, and it is possible that this concentration has an 
extension out of the surveyed area and the number of member galaxies 
might be larger than currently identified.

Daddi et al. (\cite{dad03}) examined a clustering properties of galaxies 
with photometric redshift estimates in the catalog by L03, and 
found a concentration of galaxies with red $J-K$ color ($>1.7$) 
at $z_{\mathrm phot} \sim 3.0$. If we take account the possible systematic 
offset of their photometric redshift at $z > 2$ (see \S 3.1), 
the concentration we found coincide well 
with that by Daddi et al. (\cite{dad03}), although selected galaxy 
types would be different from each other.

\section{Stellar Populations of Star-forming Galaxies at $z \sim 3$}

\subsection{Method}

The examination of stellar contents of star-forming galaxies at $z \sim 3$ 
has a crucial importance on the understanding of their star formation history 
(SFH), and their connection with galaxy populations at different redshifts. 
Deep broad-band photometric data obtained by HST/WFPC2 
(Casertano et al. \cite{cas00}) and VLT/ISAAC (L03) 
covers a rest-frame wavelength range from 
750\AA\ to 5700\AA\ for a galaxy at $z \sim 3$ with 7 bands, with which 
we can constrain some parameters of their stellar
contents, such as stellar mass, SFH and amount of 
dust attenuation, with reasonable accuracy. 
In table \ref{tab:colors} we list the six colors and $I_\mathrm{814}$
and $Ks$-band magnitudes for spectroscopic sample galaxies.

We used a population synthesis and evolution code P\'{E}GASE version 2
(Fioc and Rocca-Volmerange \cite{fio97}) for generating a set of
template galaxy spectra. 
The initial mass function adopted is that by Salpeter (\cite{sal55}) with 
a mass range of 0.1 $M_{\sun}$ -- 120 $M_{\sun}$. 
Two types of SFHs were considered, namely, 
a constant star-formation rate (SFR) and an exponentially decaying SFR.
\footnote{
For exponential SFH, four different values of the decaying timescale $\tau$
(10Myr, 50Myr, 100Myr, 1Gyr) were adopted. 
In order to model different chemical enrichment histories, 
three different SFRs 
relative to the total amount of baryons within a galaxy 
(consuming $5\times10^{-4}$, $5\times10^{-5}$, or $5\times10^{-6}$ of 
total baryons per Myr) were considered in constant SFR models.} 
P\'{E}GASE version 2 takes chemical evolution into account. 
We started calculations of synthetic spectra of stellar populations 
from zero-metal gas. The metallicity of interstellar gas of model galaxies 
at fitted ages is sub-solar. 
Although the measurements of the metallicity of LBGs still
have large uncertainties, roughly a half Solar metallicity has been
suggested (Pettini et al. \cite{pettini01}). Thus we think metallicity 
of our template spectra does not conflict with observations. 
We also calculated spectra of galaxies forming from solar initial 
metallicity, in order to examine the effect of metallicity on 
derived parameters of stellar populations. 
Stellar mass, ages and dust attenuation obtained using such high 
metallicity models did not change so much from results with 
low metallicity models; ages were a few dex younger and 
$E(B-V)$ was 0.1--0.2 smaller, but stellar mass did not change 
significantly. So we think metallicity is not a dominant source 
of uncertainty in the estimation of stellar populations. 

The procedures for finding the best-fitted models and estimating 
confidence levels we adopted are same as those made by previous 
authors (e.g., Papovich et al. \cite{pap01}). 
The outputs at 26 time steps from 10 Myr to 2.5 Gyr were used for fitting 
with observed SEDs of 13 sample galaxies (11 with firm and 2 with probable 
redshifts). Before executing the fitting, the absorption 
by intergalactic neutral hydrogen for each redshift of the object was 
applied, using analytic formula in Inoue et al. (\cite{ino05}). 
We also varied the amount of attenuation caused by dust within galaxies, 
from $E(B-V)=0.0$ to 1.5 with a step of 0.01, by adopting 
an attenuation law proposed by Calzetti et al. (\cite{cal00}). 
For each template SEDs the normalization factor which minimize 
the $\chi^2$ static were explored. The $\chi^2$ is defined as 

\begin{equation}
 \chi^2 = \sum_{i}\frac{[f_\mathrm{obs}(i) - a \times f_\mathrm{model}(i)]^2}{\sigma(i)^2},
\end{equation}
where $f_\mathrm{obs}(i)$, $f_\mathrm{model}(i)$ and $\sigma(i)$ are 
observed flux, model flux and estimated observational error in band $i$, respectively, 
and $a$ is the normalization factor. 
As observational errors we only included photometric errors listed 
in the catalogs. We also tested cases which incorporated additional errors 
possibly associated with systematic differences 
between observing bands. Although the resulting $\chi^2$ 
values were reduced in the latter procedure, the changes in best-fitting parameters 
and confidence levels were insignificant. 
For objects with Ly$\alpha$ emission line, flux densities attributed 
to Ly$\alpha$ line were estimated from our spectra and they were 
removed from $B_{450}$-band and / or $V_{606}$-band flux densities. 
For each galaxy we recorded the $\chi^2$ values calculated for all models, 
and found 
the best-fit model 
which returned the minimum $\chi^2$ value. 

The 68\% confidence levels were calculated by a Monte Carlo resampling. 
We randomly perturbed the observed fluxes, assuming the gaussian 
distribution of errors. 
The amount of the perturbations were adjusted to match the photometric errors. 
We executed the SED fitting for the set of perturbed flux densities, and the resulting 
SED parameters were recorded. The test was repeated 1,000 times for each of 
sample galaxy. Then $\chi^2$ value to contain 68\% of these simulated data 
were derived. 
This value was used to determine the range of model parameters 
for each object with 68\% confidence level, by selecting models 
which return $\chi^2$ values smaller than that for unperturbed 
object fluxes.
We also tested another way of the confidence level estimation 
with fixed $\chi^2$ values (i.e., without Monte Carlo resampling) 
following Avni (\cite{av76}). The area sizes covered by 68\% confidence 
levels are smaller than those through Monte Carlo resampling in 
most cases, while for fainter objects the area sizes are comparable. 
Hereafter we use the 68\% confidence level areas estimated 
with Monte Carlo resampling as typical errors in our SED fitting.

\subsection{Results}

\begin{figure*}
\centering
\resizebox{0.45\hsize}{!}{\includegraphics{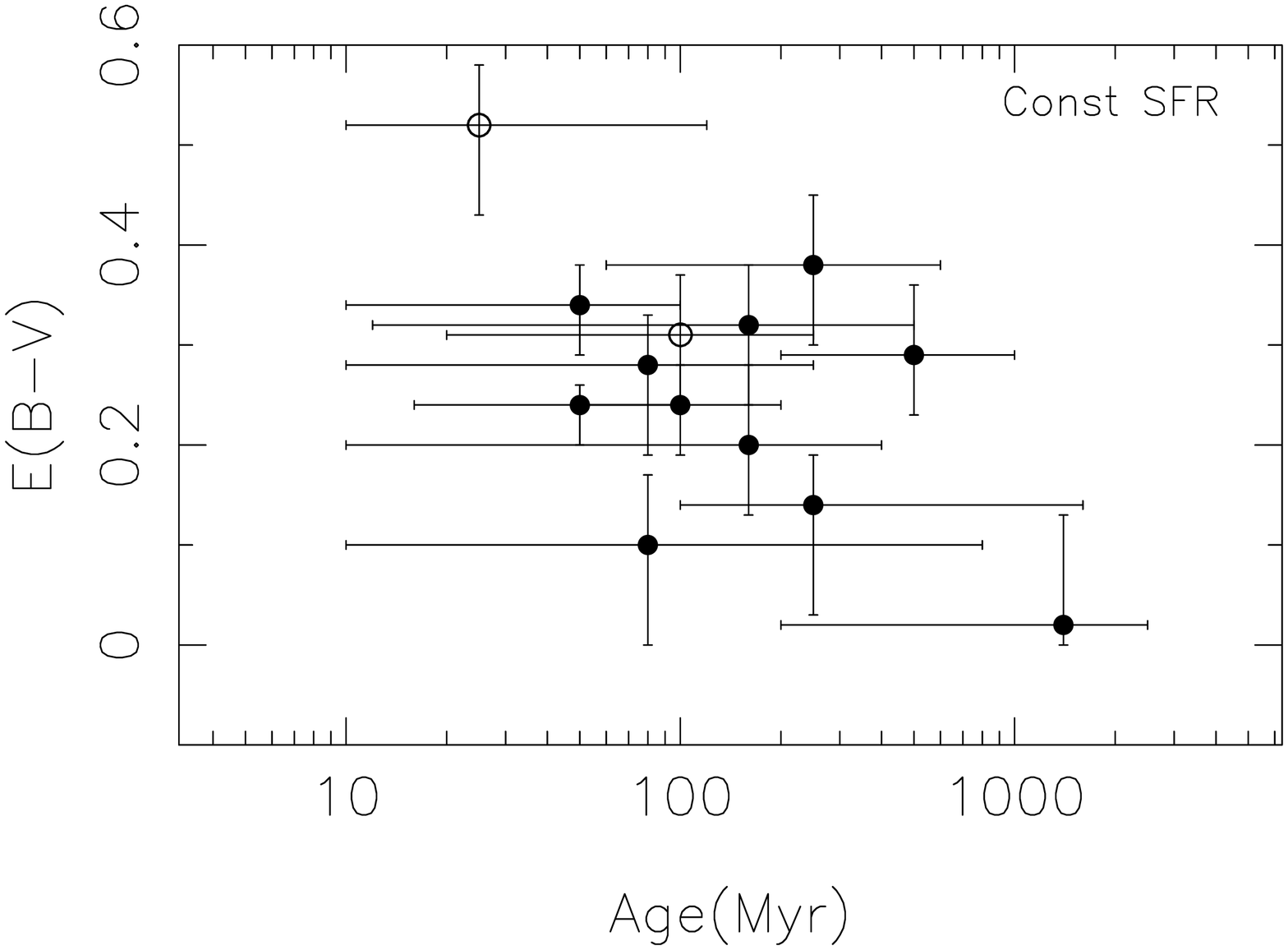}}
\hspace{2em}
\resizebox{0.45\hsize}{!}{\includegraphics{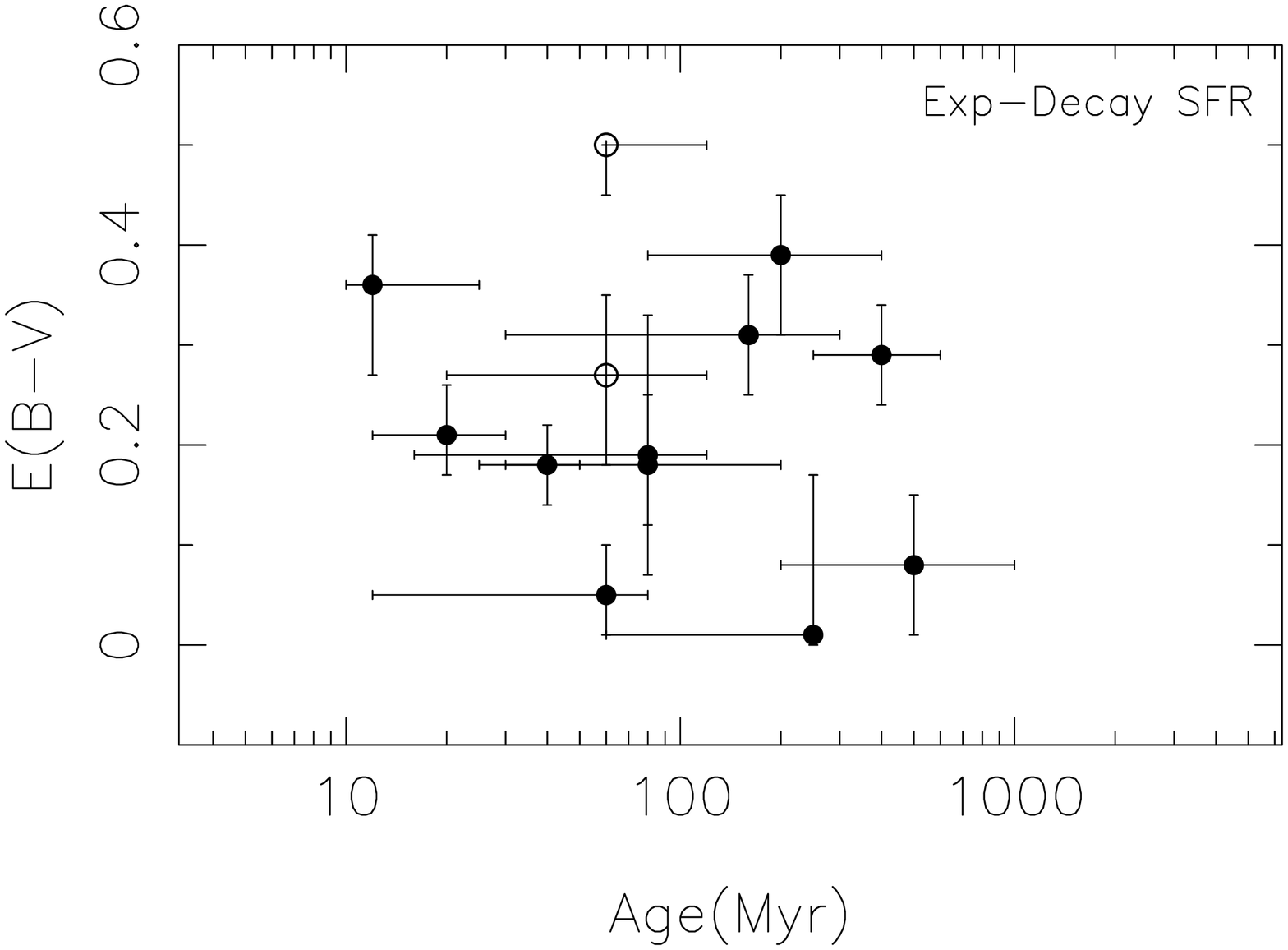}}
\caption{Distribution of ages and dust attenuation estimates obtained as results of 
SED fitting for star-forming galaxies in the HDF-S. The model spectra of 
galaxies used for comparison were calculated assuming the constant SFRs 
{\it (left)} and exponentially-decaying SFRs ({\it right}). 
Galaxies with probable redshifts (HDFS01295 and HDFS0001) are shown as open circles. 
The error bars show the range of 68\% confidence levels.}
\label{fig:sedfit1}
\end{figure*}

\begin{figure*}
\centering
\resizebox{0.45\hsize}{!}{\includegraphics{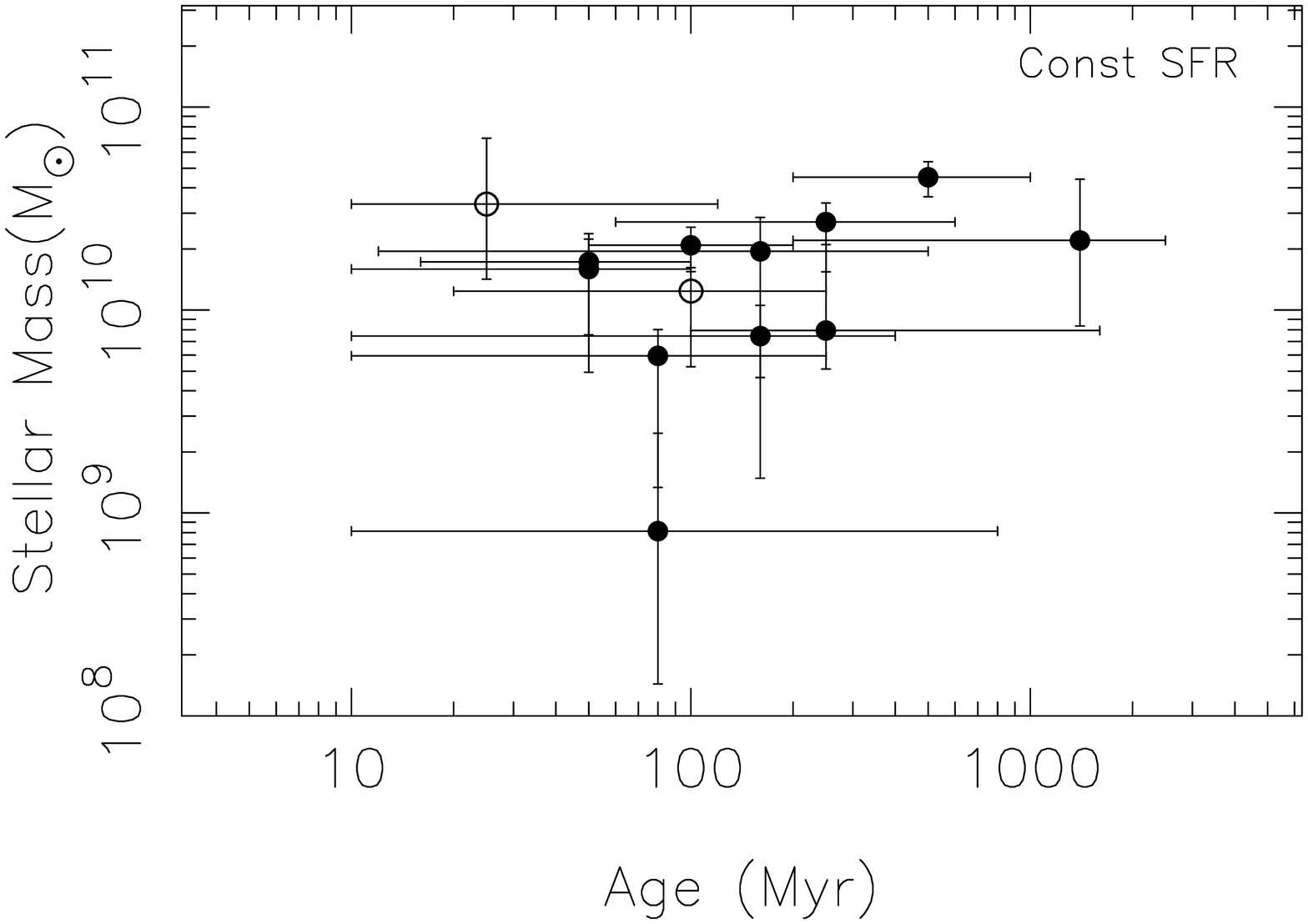}}
\hspace{2em}
\resizebox{0.45\hsize}{!}{\includegraphics{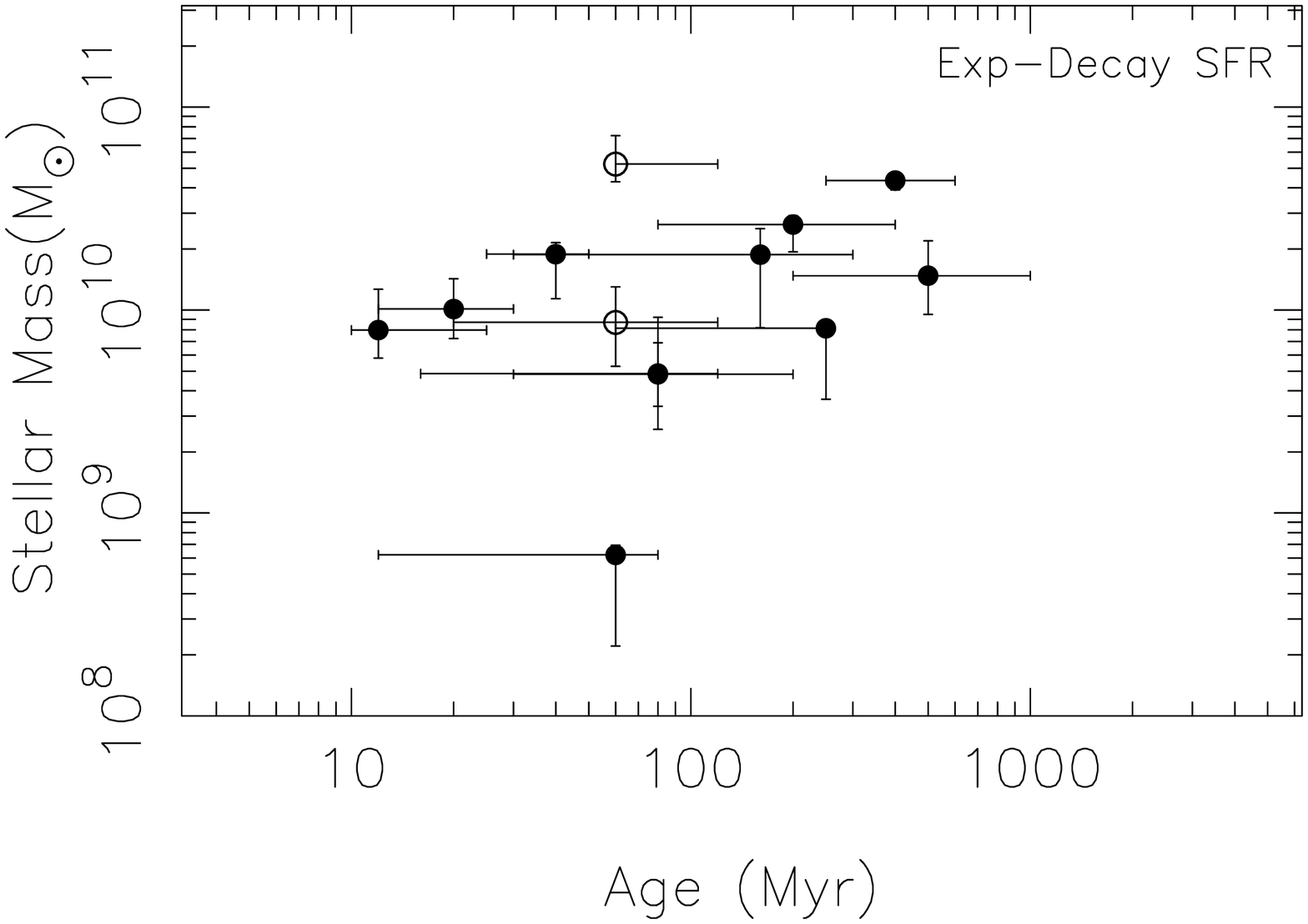}}
\caption{Ages and stellar mass estimates for star-forming 
galaxies in the HDF-S.  The meanings of symbols and error bars 
are same as Fig.~\ref{fig:sedfit1}.} 
\label{fig:sedfit2}
\end{figure*}

In Fig.~\ref{fig:sedfit1} and \ref{fig:sedfit2} we show the results of SED 
fitting. In table \ref{tab:sed1} we summarized results in case of constant 
SFRs and initial metal-free gas.
Fig. ~\ref{fig:sedfit1} shows the age from the onset of the star
formation activity and the amount of dust attenuation (in unit of color
excess $E(B-V)$). The choice of SFH (constant or 
exponentially decaying) does not make a large difference in
estimated ages and amount of dust attenuation for most cases. 
The median age and color excess of sample galaxies are 130 Myr 
and $E(B-V)=0.29$ for constant SFH, and 80 Myr and $E(B-V)=0.27$ for
exponentially-decaying SFH. When using exponentially-decaying 
SFHs, the best-fit decaying time scales are comparable to the 
age estimates. 
The median values of dust attenuation we derived are 
slightly larger than the value in some of previous studies 
($E(B-V)\sim 0.15$; e.g., Shapley et al. \cite{shap01}; 
Steidel et al.\cite{ste99}; 
but our values are consistent with those by Sawicki and Yee \cite{sy98}). 
We think the relatively smaller amount of dust attenuation in 
LBGs studied by Steidel and co-workers are due to their choice 
of filter set and color criteria. Their LBG selection criteria in $U_n-G$ and 
$G-{\cal R}$ color diagram preferentially select galaxies with $E(B-V)<0.3$ 
(see figure 2 in Steidel et al. \cite{ste03}). 
The estimated values of $E(B-V)$ and the distribution of 
our sample galaxies in two-color diagram are consistent 
with each other (see figure \ref{fig:twocol}). 

In Fig. ~\ref{fig:sedfit2} the stellar mass estimates are plotted along the 
estimated ages. The difference of stellar mass due to the choice of SFH 
is again small: the median value of stellar mass is 
$1.9\times10^{10} M_{\sun}$ and $1.5 \times 10^{10} M_{\sun}$ for
constant SFH and exponentially-decaying SFH, respectively. 
The one object which has $\sim 10$ times smaller stellar mass compared
to other objects is HDFS0271. Because the $Ks$-band magnitude of the
objects is about 2 magnitude fainter than the average of $Ks$-band
magnitude of other objects, smaller stellar mass estimate for this 
galaxy is reasonable.

The properties of stellar populations in our sample galaxies obtained by 
the SED fitting analyses are broadly 
consistent with the results of previous studies for LBGs at $z \sim 3$ 
(e.g., Sawicki and Yee \cite{sy98}; Shapley et al. \cite{shap01}; 
Papovich et al. \cite{pap01}). 

A critical argument over the estimates of stellar populations in high-redshift 
galaxies through the SED fitting technique is that there might be a
missing old stellar population which is not dominant in the wavelength
range used for the fitting but contains significant mass of low-mass stars. 
In order to explore such possibility, 
we also tested a SED fitting with model
spectra which incorporate two components of stellar populations. 
One is a single-age stellar population burstly formed in 2 Gyr ago 
($z \sim 20$ for an object at $z = 3$), 
and the other is a stellar population formed by a constant star
formation. The age and SFR of the latter were changed in
a same way as we did in the fitting with a single component. The ratio of 
total baryonic mass (including both stellar and gaseous mass) for 
old and constant components within a model galaxy were changed from
1:100 to 10:1. By this test we intended to estimate the maximal amount
of old stellar component which reasonably agrees with observations. 
For about half sample galaxies the best-fitted models were those with 
less than 10\% of the total stellar mass was coming from old component. 
For three cases (namely HDFS 0085, 1590 and 0001) which 
have stellar mass of the old component larger than those from constant SF 
component, the total stellar mass becomes 3--5 times larger than the results 
with single component fitting. 
From this test we confirmed the results by Papovich et al. (\cite{pap01}); 
the amount of old stellar population laying under the young, UV-bright 
stars seems to be not so large in the best-fitted models. 
Barmby et al. (\cite{bar04}) reported the first result of SED fitting 
for LBGs using mid-infrared observation executed by IRAC on borad the
Spitzer Space Telescope, in addition to optical and near-infrared
photometry. They argue that the stellar mass and age estimates made with
and without mid-infrared (rest-frame near-infrared) do not differ
significantly. This result also supports the idea that LBGs at $z \sim
3$ have relatively small amount of old stellar population. 
Shapley et al. (\cite{shap05}) confirmed the insensitivity of best-fit 
stellar mass estimates to the inclusion of rest-frame near-infrared 
photometry for a sample of UV-luminous star-forming galaxies at $z \sim 2$. 
However, they argued that, by considering the maximal amount of old stellar 
components, the upper limit to stellar mass for some of star-forming galaxies 
can be $>5$ times larger than the best-fitted values with the constant star 
formation models. 
In our SED fitting results we explored the 68\% confidence levels 
of stellar mass constraints in the same way as \S 4.1. 
The stellar mass of the old stellar components 
can be 2--20 times larger than the best-fitted ones, and in five 
cases old stellar components can dominate $>90$\% of the total stellar 
mass. Thus although from best-fitted models the amount of old stellar 
component is suggested to be small in most cases, 
our constraints on mass contribution from the old component are not 
strong.

We should also note that the error in estimating dust attenuation is 
another possible source of uncertainty in SED fitting. 
The best way of estimating dust attenuation would be through the 
flux ratio of far-infrared emission to UV radiation 
(e.g. Witt \& Gordon \cite{witt00}; Buat et al. \cite{buat05}) 
even for UV-selected samples like LBGs. 
Burgarella et al. (\cite{burg05}) performed a SED analysis of nearby galaxies 
selected by UV or FIR fluxes in Buat et al. (\cite{buat05}). They show that 
the amount of dust attenuation evaluated without the far infrared 
information can be wrong by up to 1 magnitude for UV-selected 
galaxies and up to 3 magnitudes for FIR-selected galaxies. 
The uncertainty of 1 magnitude on the UV attenuation leads to 
the uncertainty of 0.1 magnitude on $E(B-V)$ for the Calzetti's law.
Moreover, it seems that the dust attenuation law varies in the slope 
and in the strength of the 2175\AA\ bump from a galaxy to a galaxy.
It would enlarge the uncertainty of $E(B-V)$. Such a variation of the 
attenuation law depending on the galaxy type was examined 
by Inoue (\cite{ino05b}) theoretically. He suggest that the Calzetti's law 
is realized in a large density medium which is favorable to 
starburst galaxies like LBGs. If it is true, the assumption of 
the Calzetti's law is valid, although we need more investigations 
in this topic, especially with the data of the dust infrared emission.

Shapley et al. (\cite{shap01}) examined the correlation between the 
properties of stellar population and spectroscopic features 
for LBGs at $z \sim 3$. 
They found that galaxies with {\it young} age ($t_{sf}$, 
best fit age from the onset of the star formation, 
is shorter than or equal to 35 Myr) preferentially show 
broad Ly$\alpha$ absorption line and strong low-ionization interstellar 
lines (such as Si~{\sc ii} and C~{\sc ii}), while {\it old} 
($t_{sf} \geq$ 1 Gyr) have strong Ly$\alpha$ emission lines and weaker 
interstellar lines. 
We examined the relationship between spectroscopic features and 
parameters of best-fitted stellar population models in our 13 spectroscopic 
sample. We did not find any clear correlation between the Ly$\alpha$ emission, 
Ly$\alpha$ equivalent widths and the results of SED fitting. 
The absence of any clear trend may be attributed to the small number 
of sample galaxies in our study; the number of galaxies with spectroscopic 
identifications in Shapley et al. (\cite{shap01}) is 81, which is 
more than six times larger than ours.

\section{Discussion: Comparison with distant red galaxies at $z \sim 3$}

Franx et al. (\cite{frx03}) used deep near-infrared images obtained 
by FIRES to select galaxies which have photometric redshifts 
larger than 2 and have $Js - Ks$ larger than 2.3 (Vega) in the 
HDF-S. 
They claimed that the red $Js - Ks$ colors of these galaxies 
might be caused by the 4000 \AA\ break and bright $K$-band color 
would indicate the existence of massive stellar population. 
They also roughly estimated the number density of these ``distant 
red galaxies'' (hereafter DRGs) and suggested that they might be 
as abundant as half the number density of LBGs. 

\begin{figure*}
\centering
\resizebox{\hsize}{!}{\includegraphics{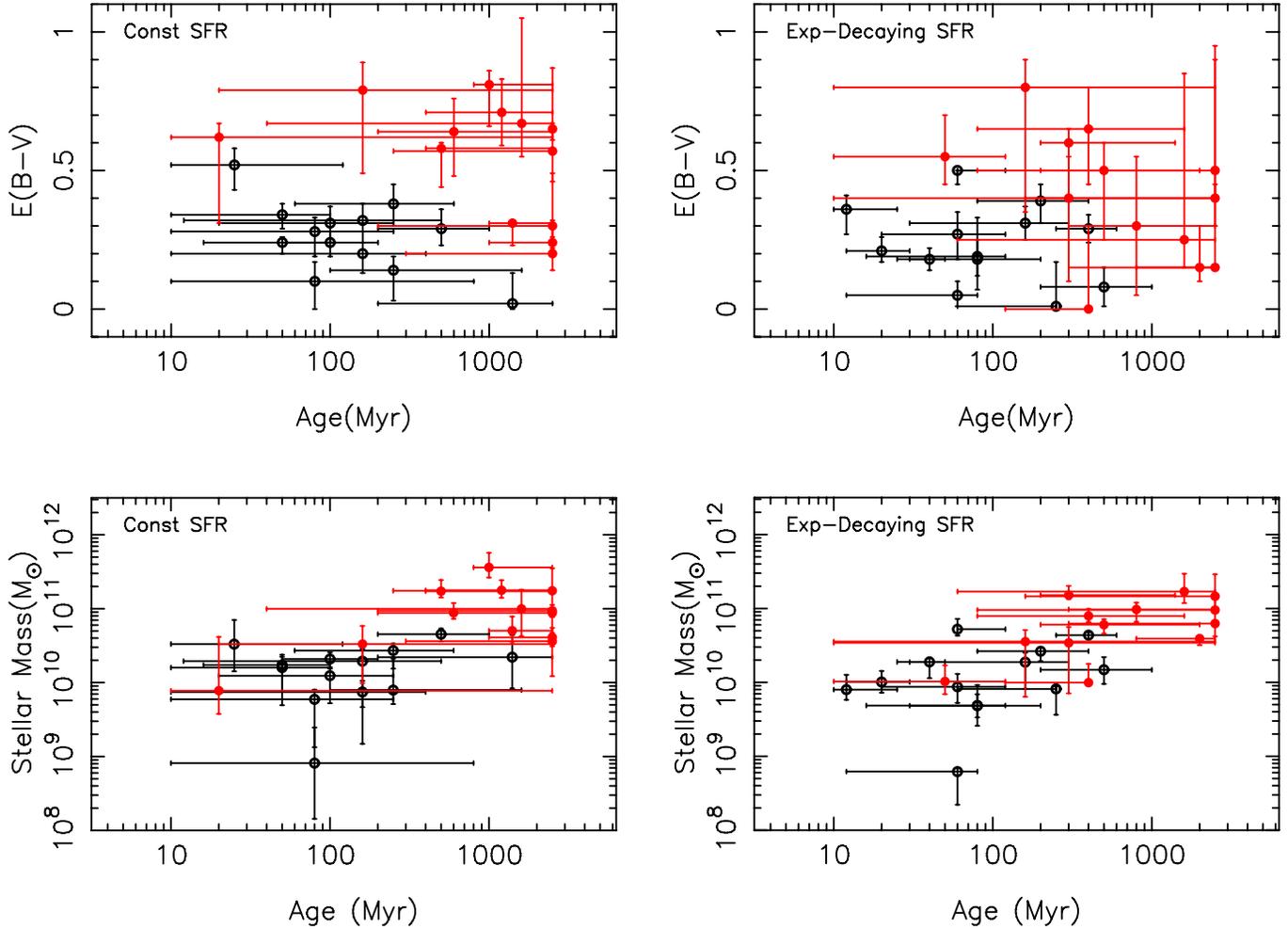}}
\hspace{2em}
\caption{Distribution of ages, dust attenuation and stellar mass 
estimates obtained as results of 
SED fitting for star-forming galaxies (open circles) and 
distant red galaxies (red filled circles) in the HDF-S. 
In the left two panels, the models assume the constant SFH 
and in the right panels the models with exponentially-decaying SFH 
were used for comparison with observed SEDs. All calculations of stellar populations 
in models in these diagrams were started from metal-free gas. 
The error bars indicate the ranges covered by 68\% confidence levels.
}
\label{fig:sedfit_drg1}
\end{figure*}

We compared the properties of stellar population 
of our sample of star-forming galaxies and DRGs. For this 
purpose we applied a SED fitting procedure for the 
13 DRGs with $K < 24.5$ in the HDF-S. The method is exactly same as that
we called ``one-component fitting'' in section 4, 
except that redshifts we used for DRGs were photometric ones derived 
by L03.
As we noted in section 3.1, 
there might be systematic offsets in photometric redshift estimates 
by L03. 
To examine the effect of this 
redshift uncertainty in analyses of stellar contents, we executed 
a test in which we adopted redshifts $-0.25$ smaller than phot-z estimates. 
The estimated amount of dust attenuation did not change significantly and 
for stellar masses, SFRs and ages differences were less 
than 50\%, mostly smaller than 30\%.
We also tested fitting with ``two-components'', i.e., 
a maximally old stellar population plus that with 
constant SFRs, and resulting parameters such as 
stellar masses and dust attenuation in young component were not 
so different from those by one-component fitting.
Both constant and exponentially-decaying 
SFHs with various timescales were tested. 
We show estimated ages (from the onset of star formation), 
dust attenuation and stellar mass in figure \ref{fig:sedfit_drg1}. 
As we did for LBG sample, we tested models with solar initial 
gas metallicity as well as those formed from metal-free gas, 
and found that differences in obtained parameters were 
smaller than 68\% confidence ranges. 
As seen in the figure, the ages of DRGs are hardly constrained. 
The 68\% confidence range of estimated ages for some objects even 
span over an order of magnitude.
Such less accurate age estimates for DRGs as compared with those for 
LBGs would be attributed to DRGs' optical faintness. Also DRGs are 
relatively abundant in low-mass stellar population, which is 
less sensible for ages. 
Given the large uncertainty of age estimation, 
the best-fitted ages of DRGs are on average larger than those of LBGs. 
The median age of our DRG sample is 1.6 Gyr, several times longer than 
that of LBGs. 
The amount of dust attenuation is also systematically larger for DRGs. 
Although we mentioned before (section 4.2), the use of single attenuation 
law may lead the error of $\sim$0.1 mag in the estimation of $E(B-V)$, 
for LBGs best fitted values of $E(B-V)$ are always not larger than 
0.5, while for DRGs $E(B-V)$ often exceeds 0.5. So a systematic difference between 
$E(B-V)$ for LBGs and DRGs seen in figure \ref{fig:sedfit_drg1} would be real.

The errors of estimated stellar mass derived from Monte-Carlo resampling 
were smaller than 0.3 dex, regardless of large uncertainty of age estimates. 
It is because stellar mass is largely dependent on the fluxes in the 
near-infrared (rest-frame optical) wavelengths. The stellar mass of 
DRGs are $10^{10}$ -- $10^{12}$ $M_{\sun}$, and the median value is 
$9.4 \times 10^{10} M_{\sun}$ for the constant SFR models. 
DRGs have systematically larger stellar masses than our sample LBGs.
We should beware that, unlike the DRG sample, our LBG sample is not complete in 
$K$-band magnitude. Thus with current data set we cannot exclude 
the possible existence of LBGs which are missed in our sample due to 
the faintness in rest-frame UV and have stellar mass comparable to DRGs. 
On the other hand, SFRs of DRGs are fairly large and 
comparable to those of LBGs: in the case of constant SFR models, 
all best-fit models have a few tens to a few hundreds $M_{\sun}$ 
per year. In the case of exponentially-decaying SFH, 7 among 13 
have on-going SFRs larger than 10 $M_{\sun}$ per year and three among 
the remaining six objects have $1 < $ SFR $<10$ 
in units of $M_{\sun}$ per year.

These results of SED fitting imply that many of DRGs are {\it not} 
passively evolving galaxies. 
DRGs would be more likely dusty galaxies with relatively 
large SFRs, and they have already accumulated 
stellar mass more than $10^{10} M_{\sun}$. 
These results of SED fitting for DRGs are consistent with some recent 
studies by other investigators. 
F\"{o}rster Schreiber et al. (\cite{fs04}) used a sample of DRGs from 
MS 1054$-$03 field as well as HDF-S for SED fitting using 
template spectra generated with a population synthesis code by 
Bruzual and Charlot (\cite{bc03}). The parameters of stellar populations 
they obtained agree well with ours.

If the number density of DRGs is as large as a half of LBGs' number density, 
the stellar mass density at $z \sim 3$ based on UV-selected galaxies 
would be significantly underestimated. Indeed, predictions based on 
cosmological numerical simulations have suggested the stellar mass 
density at redshifts larger than 2 is larger than observational 
estimations using UV-selected sample (e.g., Nagamine et al. \cite{nag04}). 
However, because near-infrared surveys deep enough to detect DRGs are 
still covers only a tiny area of the sky (smaller than several tens of 
square arcmin), uncertainty due to the cosmic variance would be large. 
In addition, they are faint in optical wavelengths and spectroscopic follow-up 
observations are much difficult than LBGs. 
Upcoming wide-field near-infrared instruments would 
much broaden survey area and increase the number of spectroscopically 
identified DRGs, which would be indispensable to derive 
reliable number density of DRGs.

\section{Summary and Conclusions}

We made an optical spectroscopy of star-forming galaxies in the 
Hubble Deep Field - South with photometric redshift estimated 
to be at $z \sim 3$. Five firm redshifts and two probable 
redshifts have been newly determined 
in addition to the confirmations of six galaxies with previously 
known redshifts. Six objects among them lie within 
$\sim$30 Mpc$^3$ comoving volume, suggesting the existence of a 
galaxy concentration.

We investigated stellar populations of sample galaxies 
using optical to near-infrared broad-band imaging data. 
By the comparisons with template spectra we obtained 
best-fit parameters. Typical ages are 20--200 Myr, 
and stellar mass is (0.5--5) $\times 10^{10} M_{\sun}$. 
These results are consistent 
with previous studies on stellar populations of LBGs, 
and we confirmed that stellar content of 
UV selected star-forming galaxies are dominated by young, 
massive stars and dust attenuation is smaller than $E(B-V)<0.5$. 
We found no significant correlation between properties of 
stellar populations and UV spectral features, in our 
11 confirmed and 2 possible sample galaxies.

A comparison of stellar contents in LBGs and DRGs in the HDF-S were 
also made. 
We found that DRGs tend to have larger values than our UV-luminous 
LBGs for stellar masses and dust attenuation, and that 
SFRs in many of DRGs and LBGs are comparable. 
Majority of DRGs would be one of the most massive, actively 
star-forming populations at $z \gtrsim 2$. 
If the number density of DRGs is comparable with that of LBGs, 
current estimates of cosmic star formation rates and stellar mass 
density at $z \sim 3$ based on UV-selected sample would be largely underestimated. 
To reach a rigid conclusion of the comparison between LBGs and DRGs 
and to draw a complete picture of the galaxy evolution in a high-redshift, 
we need deep and wide surveys in near-infrared wavelengths. 

\begin{acknowledgements}
We thank Veronique Buat, Jean-Michel Deharveng and Tsutomu Takeuchi 
for valuable discussions and encouragements.
II was supported by a Research 
Fellowship of the Japan Society for the Promotion 
of Science for Young Scientists.
AKI is supported by the JSPS Postdoctral Fellowships 
for Research Abroad.
In the first phase of this work, AKI was invited to the Laboratoire 
d'Astrophysique de Marseille and was financially supported by 
the Observatoire Astronomique de Marseille-Provence. 
Finally, we wish to thank the anonymous referee who gave us 
valuable comments.
\end{acknowledgements}

\input{hdfs_spec_bib.tex}
\input{table1.tex}

\input{table2.tex}

\input{tab_sed1.tex}

\end{document}

%% file: table1.tex
\small

\begin{table*}[!ht]
\centering
\begin{tabular}{llllllll}
\hline
\hline
   &             &                  &                  &     &    & \multicolumn{2}{c}{Ly$\alpha$} \\
\cline{7-8}
ID & Designation & $I_\mathrm{814}$ & $K_\mathrm{tot}$ & $z_\mathrm{ph}$ & $z_\mathrm{sp}$ & FWHM & EW \\ 
\hline
00565 & HDFS J223249.40$-$603226.3 & 22.928 & 22.679 & 3.00 & 2.789            & 4.3 & 10.5 \\ 
00526 & HDFS J223249.18$-$603226.0 & 23.293 & 22.706 & 3.00 & 2.789            & 4.3 & 10.5 \\ 
01655 & HDFS J223255.75$-$603333.8 & 23.488 & 23.590 & 3.40 & --               & --  & --   \\ 
00059 & HDFS J223256.38$-$603144.3 & 23.953 & 23.450 & 3.62 & 2.945$\ast$      & abs & abs  \\ 
00085 & HDFS J223246.91$-$603146.9 & 24.064 & 23.192 & 3.30 & 3.170$\dagger$   & 6.0 & 29.9 \\ 
01590 & HDFS J223304.90$-$603330.7 & 24.156 & 23.340 & 2.82 & 2.364            & abs & abs  \\ 
00153 & HDFS J223248.86$-$603154.3 & 24.349 & 22.538 & 3.00 & 2.794            & abs & abs  \\ 
01295 & HDFS J223253.43$-$603312.2 & 24.542 & 23.464 & 3.14 & (2.798)          & abs & abs  \\ 
02376 & HDFS J223254.17$-$603409.1 & 24.594 & 23.749 & 2.70 & 2.408$\ast$      & 4.8 & 10.0 \\ 
00066 & HDFS J223249.26$-$603145.0 & 24.607 & 23.076 & 3.14 & 2.801$\ast$      & abs & abs  \\ 
01983 & HDFS J223254.25$-$603350.6 & 24.616 & 23.862 & 3.62 & --               & --  & --   \\ 
01825 & HDFS J223252.03$-$603342.6 & 24.643 & 23.800 & 3.42 & 3.275$\ast$      & 3.3 &  5.9 \\ 
00704 & HDFS J223248.37$-$603237.9 & 24.737 & 22.780 & 3.00 & --               & --  & --   \\ 
00001 & HDFS J223305.01$-$603427.8 & 24.760 & 21.916 & 4.34 & (2.472)$\dagger$ & 5.7 & 43.2 \\ 
00515 & HDFS J223252.02$-$603224.1 & 24.895 & 23.318 & 3.24 & --               & --  & --   \\ 
01298 & HDFS J223251.44$-$603314.9 & 25.001 & 24.004 & 2.90 & --               & --  & --   \\ 
00708 & HDFS J223252.22$-$603237.2 & 25.167 & 23.771 & 2.54 & --               & --  & --   \\ 
01512 & HDFS J223253.12$-$603320.3 & 25.228 & 22.999 & 3.00 & 2.670$\ast$      & abs & abs  \\ 
00591 & HDFS J223254.13$-$603230.0 & 25.255 & 24.479 & 3.08 & --               & --  & --   \\ 
01411 & HDFS J223245.41$-$603320.3 & 25.270 & 24.614 & 3.58 & --               & --  & --   \\ 
02649 & HDFS J223303.59$-$603426.5 & 25.348 & 23.835 & 3.00 & --               & --  & --   \\ 
00271 & HDFS J223249.47$-$603202.1 & 25.449 & 25.432 & 2.92 & 2.806$\dagger$   & 3.9 & 26.2 \\ 
00326 & HDFS J223250.11$-$603206.6 & 25.530 & 24.731 & 2.74 & --               & --  & --   \\ 
02032 & HDFS J223303.01$-$603353.7 & 25.559 & 22.927 & 2.50 & --               & --  & --   \\ 
10570 & HDFS J223305.43$-$603342.7 & 26.017 & 24.225 & 3.26 & --               & --  & --   \\ 
\hline
\end{tabular}
\caption{Summary of results of spectroscopic observations. 
The running IDs and designations are taken from 
the version 2 of HST/WFPC2 catalog of the HDF-S 
(Casertano et al. \cite{cas00}). 
$I_\mathrm{814}$ is a corrected isophotal magnitudes in F814W filter, 
also from Casertano et al. (\cite{cas00}), 
and $Ks$-band magnitudes are 'total' magnitudes (see Labb\'{e} et al. \cite{lab03} (L03)). 
Photometric redshifts ($z_\mathrm{ph}$) are from the catalog made by L03. 
The spectroscopic redshifts obtained by our observations are also listed 
($z_\mathrm{sp}$). Asterisks denote that the redshifts were determined by our observations 
for the first time. The redshifts with $\dagger$ were based solely on the central wavelengths 
of Ly$\alpha$ lines, and two redshifts embraced with parentheses were uncertain due to the 
poor SN of the spectra. For galaxies with Ly$\alpha$ emission lines the rest-frame 
FWHMs and equivalent widths (EW) are also listed (both units are \AA).
}
\label{tab:sample}
\end{table*}

%% file: table2.tex
\small
\begin{table*}[!ht]
\centering
\begin{tabular}{llllllllll}
\hline
\hline
ID & $I_\mathrm{814}$ & $K_\mathrm{tot}$ & $U-B$ & $B-V$ & $V-I$ & $I-K$ & $J-H$ & $H-K$ & $z_\mathrm{sp}$ \\ 
\hline
00565 & 22.928 & 22.679 &  3.677 & 0.467 & 0.196 & $-0.387$ & 0.155 & $-0.032$ & 2.789 \\ 
00526 & 23.293 & 22.706 &  3.657 & 0.523 & 0.185 & 0.121    & 0.524 & 0.007    & 2.789 \\ 
00059 & 23.953 & 23.450 &$>$2.765& 0.937 & 0.464 & 0.012    & 0.527 & 0.263    & 2.945$\ast$ \\ 
00085 & 24.064 & 23.192 &  1.541 & 0.658 & 0.168 & 0.478    & 1.134 & 0.939    & 3.170$\dagger$ \\ 
01590 & 24.156 & 23.340 &  3.211 & 0.200 & 0.112 & 0.176    & 0.516 & 0.039    & 2.364 \\ 
00153 & 24.349 & 22.538 &  4.150 & 0.565 & 0.192 & 0.725    & 1.089 & 0.181    & 2.794 \\ 
01295 & 24.542 & 23.464 &  2.733 & 0.783 & 0.272 & 0.449    & 0.813 & 0.247    & (2.798) \\ 
02376 & 24.594 & 23.749 &  2.273 & 0.403 & 0.163 & 0.345    & 0.556 & 0.077    & 2.408$\ast$ \\ 
00066 & 24.607 & 23.076 &  2.353 & 0.656 & 0.299 & 0.696    & 0.912 & 0.238    & 2.801$\ast$ \\ 
01825 & 24.643 & 23.800 &$>$2.431& 0.830 & 0.207 & 0.250    & 0.764 & 0.525    & 3.275$\ast$ \\ 
00001 & 24.760 & 21.916 &  1.608 & 0.444 & 0.662 & 1.722    & 1.071 & 0.435    & (2.472)$\dagger$ \\ 
01512 & 25.228 & 22.999 &  3.402 & 0.828 & 0.360 & 1.161    & 1.046 & 0.147    & 2.670$\ast$ \\ 
00271 & 25.449 & 25.432 &  1.994 & 0.253 & 0.171 & $-0.294$ & 0.248 & 0.415    & 2.806$\dagger$ \\ 
\hline
\end{tabular}
\caption{
Colors of galaxies with spectroscopic redshifts. 
These colors are derived from magnitude measured with $0.''7$ diameter apetures. 
$I_\mathrm{814}$ and $K_\mathrm{tot}$ are same as table \ref{tab:sample}. 
}
\label{tab:colors}
\end{table*}

%% file: tab_sed1.tex
\small
\begin{table*}[!ht]
\centering
\begin{tabular}{lllllllllllll}
\hline
\hline
   & \multicolumn{3}{c}{age (Myr)} & \multicolumn{3}{c}{$E(B-V)$} & \multicolumn{3}{c}{Stellar Mass ($M_{\sun}$)}&\multicolumn{3}{c}{SFR ($M_{\sun}$/year)}\\
\cline{2-4} \cline{5-7} \cline{8-10} \cline{11-13}
ID   & best & min & max &  best & min  & max  & best & min & max & best & min & max \\
\hline
 0565 &   50 &  16 &  100 & 0.24 & 0.20 & 0.26 & 1.73e+10 & 7.55e+09 & 2.24e+10 & 3.79e+02 & 2.34e+02 & 5.13e+02 \\
 0526 &  100 &  50 &  200 & 0.24 & 0.19 & 0.28 & 2.09e+10 & 1.62e+10 & 2.56e+10 & 2.40e+02 & 1.41e+02 & 3.72e+02 \\
 0059 &   50 &  10 &  100 & 0.34 & 0.29 & 0.38 & 1.59e+10 & 4.94e+09 & 2.38e+10 & 3.50e+02 & 1.96e+02 & 6.24e+02 \\
 0085 & 1400 & 200 & 2500 & 0.02 & 0.00 & 0.13 & 2.21e+10 & 8.34e+09 & 4.41e+10 & 2.10e+01 & 1.70e+01 & 5.50e+01 \\
 1590 &  160 &  10 &  400 & 0.20 & 0.13 & 0.28 & 7.45e+09 & 1.48e+09 & 1.06e+10 & 5.50e+01 & 2.80e+01 & 1.99e+02 \\
 0153 &  500 & 200 & 1000 & 0.29 & 0.23 & 0.36 & 4.52e+10 & 3.62e+10 & 5.38e+10 & 1.15e+02 & 6.58e+01 & 2.33e+02 \\
 1295 &  100 &  20 &  250 & 0.31 & 0.24 & 0.37 & 1.24e+10 & 5.25e+09 & 1.55e+10 & 1.42e+02 & 6.90e+01 & 2.98e+02 \\
 2376 &   80 &  10 &  250 & 0.28 & 0.19 & 0.33 & 5.95e+09 & 1.34e+09 & 8.02e+09 & 8.40e+01 & 3.35e+01 & 1.93e+02 \\
 0066 &  160 &  12 &  500 & 0.32 & 0.24 & 0.38 & 1.95e+10 & 4.65e+09 & 2.86e+10 & 1.44e+02 & 6.20e+01 & 3.78e+02 \\
 1825 &  250 & 100 & 1600 & 0.14 & 0.03 & 0.19 & 7.93e+09 & 5.12e+09 & 2.10e+10 & 3.85e+01 & 1.50e+01 & 6.70e+01 \\
 0001 &   25 &  10 &  120 & 0.52 & 0.43 & 0.58 & 3.33e+10 & 1.42e+10 & 7.03e+10 & 1.39e+03 & 5.21e+02 & 1.77e+03 \\
 1512 &  250 &  60 &  600 & 0.38 & 0.30 & 0.45 & 2.72e+10 & 1.54e+10 & 3.37e+10 & 1.32e+02 & 6.15e+01 & 3.09e+02 \\
 0271 &   80 &  10 &  800 & 0.10 & 0.00 & 0.17 & 8.14e+08 & 1.44e+08 & 2.47e+09 & 1.15e+01 & 4.00e+00 & 3.15e+01 \\
\hline
\end{tabular}
\caption{Results of SED fitting, using template SEDs with constant star
 formation rates which started with zero metallicity gas. 
 ``Min'' and ``max'' were lower and upper values for 68\% confidence levels 
 obtained through Monte-Carlo resamplings.
}
\label{tab:sed1}
\end{table*}